\newcommand{\gev}{\, {\rm GeV}}
\newcommand{\tev}{\, {\rm TeV}}

\documentclass[twocolumn,preprintnumbers,superscriptaddress]{revtex4}
\usepackage{graphicx}
\usepackage{amssymb,amsmath,bm}
\usepackage{epstopdf}
\usepackage{mathrsfs}
\usepackage{amssymb}
\usepackage{verbatim}
\usepackage{color}
\usepackage{multirow}
\usepackage{subfigure} 

\begin{document}

\preprint{LUPM:15-002}

\vspace*{1mm}

\title{Light Sparticles from a Light Singlet in Gauge Mediation}

\author{Ben~Allanach}
\affiliation{
DAMTP, CMS, Wilberforce Road, University of Cambridge, Cambridge, CB3 0WA, United Kingdom}
\author{Marcin~Badziak}
\affiliation{
Institute of Theoretical Physics, Faculty of Physics, University of Warsaw,\\
ul. Pasteura 5, PL-02-093 Warsaw, Poland}
\author{Cyril~Hugonie}
 \affiliation{
LUPM, UMR 5299, CNRS, Universit\'e de Montpellier II, 34095 Montpellier, France}
\author{Robert~Ziegler}
\affiliation{
Sorbonne Universit\'es, UPMC Univ Paris 06, UMR 7589, LPTHE, F-75005, Paris, France}
\affiliation{
CNRS, UMR 7589, LPTHE, F-75005, Paris, France}

\vspace{0.1cm}

\begin{abstract} 
We revisit a simple model that combines minimal gauge mediation and the next-to-minimal supersymmetric standard model. We show that one can obtain a
125 GeV Standard Model-like Higgs boson with stops as light as 1.1 TeV, thanks to the mixing of the Higgs with a singlet state at ${\cal O}(90-100)$ 
GeV. Sparticle searches at the LHC may come with additional $b-$jets or taus and may involve displaced vertices. The sparticle production
cross-section at the 13 TeV LHC can be ${\mathcal O}(10-100)$ fb, leading to great prospects for discovery in the early phase of LHC Run II.
\end{abstract}

\maketitle

\setcounter{equation}{0}

%%%%%%%%%%%%%%%%%%%%%%%%%%%%%%%%%%%%%%%%%%%%%%%%%%%%%%%%%%%%%%%%%%%%%%

\section{Introduction}
The discovery of a Higgs-like scalar particle with mass close to 125 GeV \cite{Higgs_discovery} has considerable impact on supersymmetric (SUSY) model
building. 
In its simplest realization, the Minimal Supersymmetric Standard Model (MSSM),
the tree-level Higgs mass is bounded from above by the Z-boson mass, which implies that
large radiative corrections of the order of the tree-level mass are needed~\cite{Ruderman}. This motivates extensions of the minimal model with new
tree-level contributions to the Higgs mass. A possible source of enhancement of the tree-level Higgs mass is mixing with an additional neutral state
that is lighter than the SM-like Higgs. This situation
can be realized in the Next-to-MSSM (NMSSM)~\cite{NMSSM}. Analyses of the generic NMSSM parameter space have already demonstrated that this
possibility is viable~\cite{mixing, BOP}. 
Here instead we want to study this scenario in a simple and predictive framework of SUSY breaking, gauge mediation~\cite{GMSB}, which solves elegantly
SUSY CP
and flavor problems.

Indeed the combination of the NMSSM and gauge mediation is particularly motivated, as the NMSSM provides a simple solution to the notorious
$\mu-B_\mu$ problem~\cite{muBmu} of gauge mediation. Yet it is very difficult to realize this scenario with minimal gauge mediation (MGM), as the
NMSSM soft terms are too small~\cite{DineNelson}. These problems can however be cured by adding direct couplings of the singlet to messengers, at the
cost of a single new parameter (for a different approach see
e.g.~Ref.~\cite{Vempati}). A viable model of this kind has been proposed by
Delgado, Giudice and Slavich (DGS) in Ref.~\cite{DGS}
(extensions with Higgs-messenger couplings were
studied in Ref.~\cite{Shih}). However, the 
authors of Ref.~\cite{DGS} concluded that in this model sparticles cannot be lighter than in MGM.

In this article we re-analyze the DGS model and identify new viable regions in the parameter space where singlet-Higgs mixing is small enough to pass
experimental constraints, but large enough to give substantial contributions to the tree-level Higgs mass. This model can therefore rely on smaller
contributions from stop loops, thus reducing the overall scale of sparticle masses. Interestingly,
squarks and gluinos can be light enough to be discovered in the early stage of the LHC run II, in contrast to MGM, where a
125 GeV Higgs mass requires colored sparticles beyond the reach of the LHC (even for very high luminosity) \cite{mgmsb}. Moreover, we find that the
light singlet-like scalar can easily explain the $2\sigma$ excess around 98 GeV observed in the LEP Higgs searches \cite{LEP, BEGJKS}. The realization
of this scenario, with maximal contribution to the tree-level Higgs mass from mixing, fixes almost all of the model parameters. A single parameter
remains free and controls the details of the phenomenology. The lightest supersymmetric particle (LSP) is the gravitino and the next-to-LSP (NLSP) is
the singlino, a setup that leads to new signatures at collider experiments. The underlying model might therefore serve as a representative for a whole
class of signatures that motivate suitable SUSY search strategies.

\section{The DGS Model}
The field content of the DGS model (see Ref.~\cite{DGS} for details) consists of the NMSSM fields (the MSSM fields plus a gauge singlet $S$), in
addition to two copies of messengers in ${\bf 5 + \bar{5}}$ of SU(5), denoted by $\Phi_i, \bar{\Phi}_i$, $i= 1,2$ with SU(2) doublet and SU(3) triplet
components $\Phi_i^D, \bar{\Phi}_i^D, \Phi_i^T, \bar{\Phi}_i^T$, $i= 1,2$. Supersymmetry breaking is parametrized by a non-dynamical background field
$X = M + F \theta^2$. Apart from the Yukawa interactions,  the superpotential is given by the NMSSM part, the spurion-messenger couplings and the
singlet-messenger couplings,  $W = W_{\rm NMSSM} + W_{\rm GM} + W_{\rm DGS}$, where
\begin{align}
W_{\rm NMSSM}  & =   \lambda S H_u H_d + \frac{\kappa}{3} S^3  \, ,  \\
\label{GM}
W_{\rm GM} & =  X \sum_{i=1,2 } \left( \kappa_i^D \bar{\Phi}_{i}^{D} \Phi_{i}^{D} + \kappa_i^T \bar{\Phi}_i^{T}\Phi_i^{T}\right) \, , \\
W_{\rm DGS} & =   S \left( \xi_D \bar{\Phi}_{1}^{D} \Phi_{2}^{D} + \xi_T  \bar{\Phi}^{T}_{1} \Phi^{T}_{2} \right) \, . 
\label{DGS}
\end{align}
The new couplings in Eq.~(\ref{DGS}) are assumed to unify at the GUT scale $\xi_D (M_{\rm GUT}) = \xi_T (M_{\rm GUT}) \equiv \xi$. A similar
assumption can be made for $\kappa_i^{D,T}$, but these parameters are largely irrelevant for the spectrum. 

${\mathcal Z}_3$ invariant NMSSM models such as this one have a potential
cosmological problem with domain walls, which are predicted to appear during
the phase transition associated with electroweak symmetry breaking. This can
be solved by the introduction of higher dimensional ${\mathcal Z}_3$-violating
operators. Dangerous tadpoles that may threaten successful electroweak
symmetry breaking may be avoided by the imposition of a discrete
${\mathcal R}-$symmetry on the higher dimension
operators~\cite{Panagiotakopoulos:1998yw}.   

Through the superpotential in Eq.~(\ref{GM}) the messengers feel SUSY breaking at tree-level and communicate it to the NMSSM fields via gauge
interactions and the direct couplings in Eq.~(\ref{DGS}). The contribution from gauge interactions is given by the usual expressions of MGM for
one-loop gaugino masses $M_i$ and two-loop sfermion masses $m_{\tilde{f}}$ at the messenger
scale $M$
\begin{align}
 M_i & =  2 g_i^2 \tilde{m} \, ,  &
  m^2_{\tilde{f}}  & =  4 \sum_{i=1}^3 C_i(f)~g_i^4  \tilde{m}^2 \, , 
\end{align}
where $\tilde{m} \equiv 1/(16 \pi^2) F/M$ and $C_i(f)$  is the quadratic Casimir of the representation of the field $f$ under ${\rm SU(3)}\times {\rm
SU(2)}\times {\rm U(1)}$. The contributions from direct singlet-messenger couplings generate one-loop A-terms for the NMSSM couplings 
\begin{equation}
A_{\lambda}  =  \frac{A_{\kappa}}{3} = - \tilde{m} \left( 2 \xi_D^2 + 3 \xi_T^2 \right)\, , 
\end{equation}
and two-loop contributions to soft masses for the singlet and the Higgs fields 
\begin{align}
\tilde{m}_{S}^2 & =    \tilde{m}^2 \left[ 8 \xi_D^4 + 15 \xi_T^4 + 12 \xi_D^2 \xi_T^2 \right] \nonumber \\
& -  \tilde{m}^2  \left[ 4 \kappa^2 \left(2 \xi_D^2 + 3 \xi_T^2 \right) \right] \nonumber \\
& -  \tilde{m}^2  \left[ \xi_D^2 (\frac{6}{5} g_1^2 + 6 g_2^2 ) + \xi_T^2 (\frac{4}{5} g_1^2 + 16 g_3^2) \right]\, ,    \\
\Delta \tilde{m}_{H_u}^2  & =   \Delta \tilde{m}_{H_d}^2 = - \tilde{m}^2 \lambda^2 \left( 2 \xi_D^2 + 3 \xi_T^2 \right)\, . 
\end{align}
There is also a one-loop contribution to the singlet soft mass~\cite{DGS} that  is relevant only for very low messenger scales and has negligible
impact on the spectrum. The model is thus determined by five parameters: $\tilde{m}$, $M$, $\lambda$, $\kappa$ and $ \xi$, where one parameter
(following DGS we choose $\kappa$) can be eliminated by requiring correct electroweak symmetry breaking (EWSB). 

\section{Low-energy Spectrum}
There are several regions in the parameter space that lead to a viable spectrum and are compatible with perturbative couplings up to the GUT scale.
They can be distinguished by the size of the relative contributions to the SM-like Higgs mass, which are given
schematically by 
\begin{equation}
\label{mh2}
m_h^2 = M_Z^2\cos^2 2\beta + \lambda^2 v^2\sin^2 2\beta  +  m_{h, \rm mix}^2 +  m_{h, \rm loop}^2 \, ,
\end{equation} 
where $m^2_{h, \rm mix}$ is the contribution from mixing with the other two CP-even states in the full Higgs mass matrix. 

Since a larger tree-level mass implies a lighter SUSY spectrum, we concentrate here on the region in parameter space where the effective tree-level
contribution to the Higgs mass is maximized. 
It turns out that one can reach up to $m_{h}^2 -m_{h,\rm loop}^2  \approx (99 \gev)^2$, provided that $\tan \beta$ is large and the contribution from
mixing is sizable and positive. This requires the singlet state to be lighter than the SM-like Higgs, which is not excluded by LEP if the mixing angle
with the SM-Higgs is small enough. Note that the LHC constrains this scenario only through
measurements of the SM-like Higgs couplings that are suppressed by the mixing. 

A focal feature of this scenario is its high level of predictivity, as three out of four free parameters of the model are determined by the Higgs
sector. Maximizing the tree-level contribution to the SM-like Higgs mass $m_{h_2}$ fixes the singlet-like Higgs mass $m_{h_1}$ and the singlet-Higgs
mixing angle $\theta$ (mixing with the heavy Higgs doublet is negligible) along the lines of Ref.~\cite{BOP}, giving approximately $m_{h_1} \approx 94
\gev$ and $\cos \theta \approx 0.88$. This in turn determines the model parameters $\lambda$ and $\xi$, while the overall scale of soft terms
$\tilde{m}$ is fixed by the required size of the residual loop contribution to $m_{h_2}$. Departing from the maximal tree-level contribution leads to
slightly different $m_{h_1}$ and $\theta$, and of course larger $\tilde{m}$.   

In practice one can map the model parameters to the low-energy spectrum only numerically. For this analysis we have used a modified version of {\tt
NMSSMTools}~\cite{NMSSMTools}. Independent checks using a modified version of {\tt SOFTSUSY3.4.1}~\cite{SoftSUSY} produced Higgs and SUSY spectra that
agreed at the percent level.

Before discussing our results, we provide some rough analytic results that can be obtained neglecting renormalization group (RG) effects and expanding
the NMSSM vacuum conditions in the limit of large singlet vacuum expectation value. In this way one obtains approximate relations between $\xi,
\lambda$  and the physical Higgs parameters
\begin{align}
\xi & \sim \frac{m_{h_1}}{4 \sqrt{2} g_3  {\tilde{m}} }\,, &
\lambda \sim \frac{m_{h_2}^2-m_{h_1}^2}{4 v \tilde{m} } \sin 2 \theta \, .
\end{align}
For TeV-scale superpartners (and the above values for $m_{h_1}$ and $\theta$ that maximize the Higgs tree-level contribution) one finds $\xi \sim
\lambda \sim10^{-2}$.  

The smallness of $\xi$ and $\lambda$ has important consequences for the low-energy spectrum. Small $\xi$ implies small values of $|A_\lambda|$ and
$|A_\kappa|$ and imposing proper EWSB yields $\kappa\ll\lambda$ and the prediction of large $\tan\beta \sim \lambda/\kappa$. In turn, the smallness of
$A_\kappa$ and $\kappa$ results in a very light singlet-like pseudoscalar $a_1$ of mass
\begin{equation}
{m_{a_1}}\sim \sqrt{ \frac{45\sqrt{8}\xi}{32g_3} }{m_{h_1}} \,. \label{ma1}
\end{equation}
For $\xi
\sim 10^{-2} $, Eq.~(\ref{ma1}) predicts $m_{a_1}$ to be smaller than $m_{h_1}$
by a factor of a few. We find numerically that the light  pseudoscalar mass varies between 20 and 40 GeV, for gluino masses below 2.5 TeV.

For the above range of parameters the mass of the singlino can be obtained from the following approximate sum rule \cite{sumrule}:
\begin{equation}
 m^2_{\tilde{S}} \approx m_{h_1}^2 + \frac{1}{3} m_{a_1}^2 \,, \label{msing}
\end{equation}
which implies that the singlino mass is about 100 GeV. Since in MGM the LSP is the gravitino and the typical scale of the NLSP is the bino mass $M_1 
\approx 420 \gev \left(\tilde{m}/\tev \right) $,  it is clear that here the singlino strongly dominates the composition of the NLSP. This is a
distinguishing feature of this model. 

This is closely connected to the main virtue of this scenario, the
large contribution to the tree-level Higgs mass from singlet-Higgs
mixing. This requires smaller radiative corrections from stop loops, and in turn much lighter sparticle masses than in MGM. Through these corrections
the observed Higgs mass of 125 GeV
essentially fixes the overall scale of the sparticle spectrum $\tilde{m}$, up to an estimated theoretical uncertainty of 3 GeV in the prediction of
$m_{h_2}$.
\begin{figure}
\includegraphics[scale=0.35]{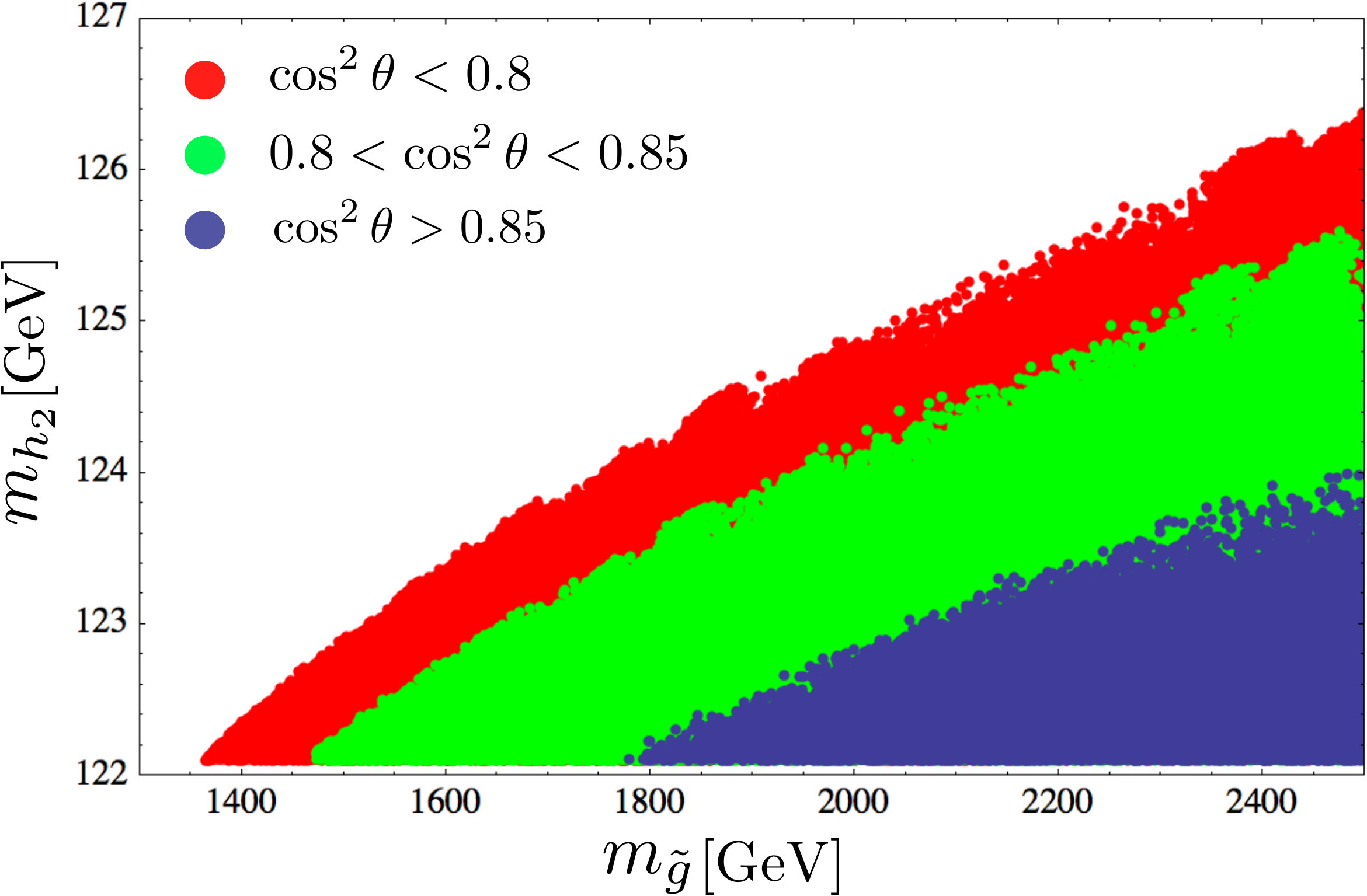}
\includegraphics[scale=0.35]{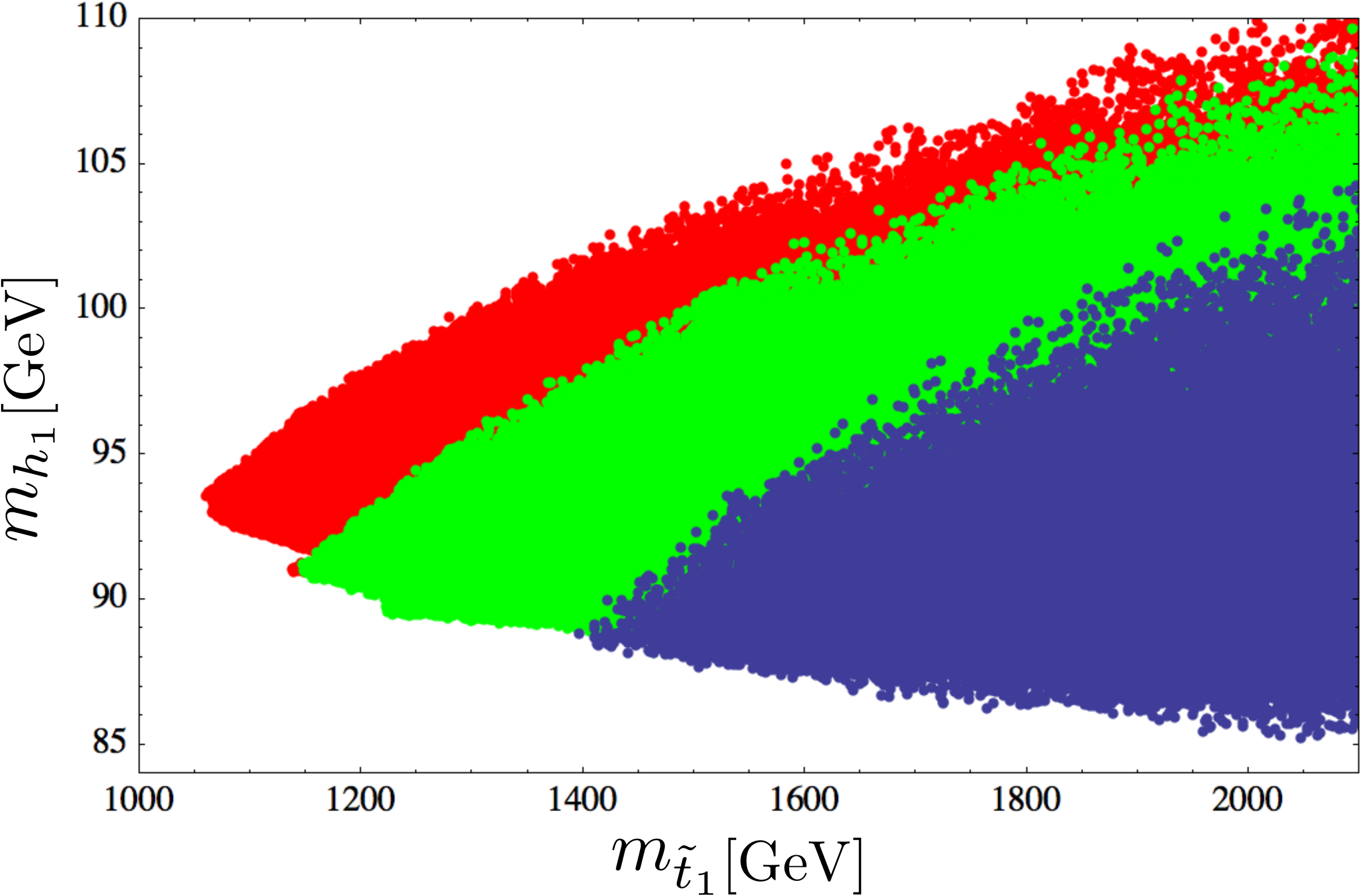}
\caption{\label{fig1} 
(Upper panel) SM-like Higgs mass vs. gluino mass and (Lower panel) singlet-like Higgs mass vs. lightest stop mass. The various  model points are
distinguished by the Higgs-singlet mixing angle $\theta$, which decreases from top to bottom as specified in the upper panel. For the same SM-like
Higgs mass a larger mixing angle allows for much lighter gluinos. The lightest stop masses are obtained for a singlet-like Higgs around 94 GeV.}
\end{figure}
We find that $m_{h_2}=125$ GeV is compatible with a gluino mass of 2.1 TeV (1.4 TeV if the theoretical uncertainty of 3 GeV for the Higgs mass is
taken into account). Squarks of the first two generations have approximately the same mass as the gluino and should be within the reach of the LHC Run
II. Stop masses can be as light as 1.7 (1.1) TeV, which should be compared with the lower bound on stop masses of about 8 (3) TeV in MGM~\cite{mgmsb}.
In Fig.~\ref{fig1} we show the values of Higgs, gluino and stop masses for several model points separated by the singlet-Higgs mixing angle $\theta$.
Note that $\cos^2 \theta$ is roughly of the size of effective Higgs signal strengths $R_i = (\sigma \times {\rm BR})_i / (\sigma \times {\rm
BR})_i^{\rm SM}$, which are substantially reduced in this scenario. Nevertheless all shown points are compatible with LEP and LHC constraints
on the Higgs sector.  

Having fixed $(\xi, \lambda, \tilde{m})$ by the set of physical Higgs parameters $( m_{h_1}, m_{h_2}, \theta)$, the only free parameter left is the
messenger scale $M$. This parameter controls the low-energy
spectrum in several ways. First of all, increasing $M$ leads to larger values
 of $A_t$ at the electroweak (EW) scale, which (as in MGM) is purely radiatively
generated and therefore grows with the length of the RG running. In turn, this enhances the stop-mixing contribution to the Higgs mass, and therefore
larger
$M$ leads to lighter stops and hence smaller $\tilde{m}$ for the same value of $m_{h_2}$. Also, the value of $M$ essentially determines the nature of
the next-to-NLSP (NNLSP). For small $M\lesssim 10^8 \gev$  the (mostly right-handed) stau is the NNLSP (with selectron and smuon being co-NNLSP),
because the soft mass $m_{\tilde{E}}$ is smaller than $M_1$ at the messenger scale. For $M\gtrsim 10^{9} \gev$ (requiring gluino masses below 2.5 TeV)
the RG effects are strong enough to raise $m_{\tilde{E}}$
above $M_1$ and the bino-like neutralino becomes the NNLSP. In the transition region $10^8 \gev \lesssim M \lesssim 10^9 \gev$ the NNLSP can be either
stau or bino, depending on the other parameters. The messenger scale $M$ controls the gravitino mass according to:
\begin{equation}
m_{3/2} = 38 \, {\rm eV} \left( \frac{\tilde{m}}{\tev} \right) \left(
  \frac{M}{10^6 \gev} \right) \, . \label{m32}
\end{equation} 
The simultaneous presence of gravitino LSP and singlino NLSP leads to a novel phenomenology  quite different both from MGM models and from typical
NMSSM scenarios.
 
\section{LHC Phenomenology}
In our scenario decay chains of every supersymmetric particle produced at the LHC end up in a singlino-like neutralino $ \tilde{N}_1$. Since the
singlet couples very weakly, these decays always proceed through the NNLSP or co-NNLSP. The singlino subsequently decays to the gravitino and the
singlet-like pseudoscalar $a_1$, which in turn predominantly decays to $b$-quarks: 
\begin{equation}
 \tilde{N}_1 \to a_1 \tilde{G} \to b \overline{b} \tilde{G} \,.
\end{equation}
The decay length of the neutralino (in its rest frame) is approximately given by 
\begin{equation}
c \tau_{\tilde{N}_1} \approx 2.5 \, {\rm cm} \, \left( \frac{100 \, {\rm GeV}}{M_{\tilde{N}_1}} \right)^5 \left( \frac{M} {10^6 \, {\rm GeV}}\right)^2
\left( \frac{\tilde{m}}{{\rm TeV}} \right)^2 \, . \label{ctau}
\end{equation}
Since $M$ cannot be much below $10^6$ GeV, it is clear from the above formula that the singlino NLSP (with mass about 100 GeV) always travels
macroscopic distance before it decays.  For large $M$ the singlino decays well outside the detector so it is stable from the collider point of view.
However, for $M\sim10^6-10^7 \gev$ the singlino may decay in the detector after traveling some distance from the interaction point leading to a
displaced vertex.
Since the value of $M$ also decides about the nature of the
NNLSP, it can be used to define three regions with distinct LHC
phenomenology, which we briefly discuss in the remainder of this letter. A more detailed analysis of LHC phenomenology and discovery prospects will be
the subject of a future publication. 

In Table \ref{tab:benchmarks} we collect several characteristic benchmark points. Points P1 and P4 represent the lightest SUSY spectra we have found,
for very low and very large messenger scales, respectively. Since the Higgs mass errors are pushed to the limits, we consider these points merely as
limiting cases, although not necessarily unrealistic. Note in particular that P4 is not obviously ruled out by standard SUSY searches for jets +
missing $E_T$, since the additional decay of the would-be-LSP bino to singlino reduces efficiency compared to the
CMSSM~\cite{sumrule,EllwangerSinglinoLSP}. The other points are representatives for the three characteristic regions discussed below, and P3 is in
addition chosen to fit the LEP excess. Note that all points have quite large singlet-Higgs mixing, leading to reduced effective Higgs couplings.
Points with smaller mixing and/or larger Higgs masses can be obtained by increasing the overall SUSY scale $\tilde{m}$.   
\begin{table}
\centering
\begin{tabular}{c|ccccc}
& {\rm P1} & {\rm P2} & {\rm P3} & {\rm P4}  &  {\rm P5}  \\
\hline
$\tilde{m}$  & $7.5 \cdot 10^2$ & $8.7 \cdot10^2$ & $9.3 \cdot10^2$ & $5.9 \cdot10^2$ &  $9.3 \cdot10^2$ \\
$M$  & $1.4 \cdot 10^{6}$ & $2.8 \cdot 10^{6}$ & $3.3 \cdot 10^{7}$ & $8.3 \cdot 10^{14}$  & $3.4 \cdot 10^{14}$ \\
$\lambda$ & $1.0  \cdot 10^{-2}$ & $9.3  \cdot 10^{-3}$ & $6.7  \cdot 10^{-3}$ & $9.2  \cdot 10^{-3}$  & $6.9  \cdot 10^{-3}$  \\
$\xi $ & $1.2  \cdot 10^{-2}$ & $1.1  \cdot 10^{-2}$ & $1.3  \cdot 10^{-2}$ & $3.2  \cdot 10^{-2}$  & $2.0  \cdot 10^{-2}$  \\
\hline
%$\kappa$ & $7.0 \cdot 10^{-4}$ & $5.4 \cdot 10^{-4}$ & $3.5 \cdot 10^{-4}$ & $5.7 \cdot 10^{-4}$  & $2.9 \cdot 10^{-4}$ \\
$\tan \beta$ & 25 & 28 & 24 & 26  & 21 \\
\hline
$m_{h_1}$ & 92 & 93 & 98 & 94  & 94 \\
$m_{h_2}$ & 122.1 & 123.4 & 122.9 & 122.1  & 125.0 \\
$m_{a_1}$ & 26 & 26 & 28 & 40  & 32 \\
$m_{\tilde{N}_1}$ & 101 & 102 & 106 & 104  & 104 \\
$m_{\tilde{N}_2}$ & 322 & 377 & 400 & 251  & 379 \\
$m_{\tilde{e}_1}$ & 303 & 358 & 406 & 449  & 676 \\
$m_{\tilde{\tau}_1}$ & 284 & 333 & 376 & 432  & 637 \\
$m_{\tilde{g}}$ & 1.73 & 1.98 & 2.09 & 1.37  & 2.06 \\
$m_{\tilde{u}_R}$ & 1.79 & 2.06 & 2.15 & 1.36  & 2.07 \\
$m_{\tilde{t}_1}$ & 1.64 & 1.87 & 1.90 & 1.06  & 1.63 \\
\hline
$c \tau_{\tilde{N}_1}$ & $6.4 \cdot 10^{-2}$ & $0.34$ & $48$ & $1.9 \cdot 10^{16}$  & $6.0 \cdot 10^{15}$ \\
\hline
$\sigma_{\tilde{q}\tilde{q}}^{13 \rm TeV}$  & 9.35 & 2.99 & 1.98 & 59.7  &  2.63 \\
$\sigma_{\tilde{q}\tilde{g}}^{13 \rm TeV}$  & 11.9 & 3.30 & 2.01 & 91.1  &  2.48 \\
%$\sigma_{\tilde{g}\tilde{g}}^{13 \rm TeV}$  & 2.19 & 0.54 &  & 18.2  & \\
%$\sigma_{\tilde{q}\tilde{\bar{q}}}^{13 \rm TeV}$  & 1.62 & 0.41 & & 16.7  & \\
%$\sigma_{\tilde{t}_1\tilde{\bar{t}}_1}^{13 \rm TeV}$  & 0.11 & 0.03 & & 3.71  & \\
$\sigma_{\rm strong}^{13 \rm TeV}$  & 25.2 & 7.28 & 4.58 & 190  &  5.95 \\
$\sigma_{\rm strong}^{8 \rm TeV}$  & 0.51 & 0.07 & 0.03 & 10.1  &  0.05 \\
$\sigma_{\rm EW}^{13 \rm TeV}$  & 27 & 12 & 7.5 & 6.7  &  5.6 \\
$\sigma_{\rm EW}^{8 \rm TeV}$  & 5.5 & 2.1 & 1.2 & 1.3  &  0.7
\end{tabular}
\caption{List of benchmark points. All masses are in GeV except colored sparticle masses in TeV, the neutralino decay length $c \tau_{\tilde{N}_1}$ in
m and cross-sections  in fb. All points have reduced effective Higgs couplings, with Higgs signal
strenghts about
$0.75$, as a result of a Higgs-singlet mixing angle with $\cos \theta \approx 0.88$.}.
\label{tab:benchmarks}
\end{table}

In all regions sparticles can be very light, so that huge parts of the parameter space are in the reach of LHC Run II.  As can be seen from Table
\ref{tab:benchmarks} the total strong production cross-section (dominated by $\tilde{q}\tilde{q}$ and  $\tilde{q}\tilde{g}$) is  ${\cal O} (10-100)$
fb, as computed with {\tt PROSPINO}~\cite{prospino}. LHC Run II is expected to deliver ${\cal O} (10)$ fb$^{-1}$ of integrated luminosity in 2015,
which results in ${\cal O} (100-1000)$ potentially
discoverable events. The total EW production cross-section at the 13 TeV LHC (computed with {\tt Pythia 8.2} \cite{Pythia}) is typically
comparable to the strong
one but is distributed among many different channels with rather small individual cross-sections of order ${\cal O} (1-10)$ fb. The most
frequent  EW
production channel is $\chi_1^+\chi_3^0$ (which are wino-like states decaying dominantly to staus) with the cross-section of about one fifth of the
total EW cross-section
\footnote{At the 8 TeV LHC the EW production cross-section is larger than the strong production cross-section (except for P4 where it is comparable).
Nevertheless, current LHC
limits for direct EW production are far too weak to constrain the model. The dominant EW production channel is a production of wino-like
charginos and neutralinos with masses of about 600-700 GeV (except for P4), which subsequently decay dominantly to staus. The lower mass limits for
charginos decaying into staus has been set in some simplified models but are always below 400 GeV
\cite{chargino_viastau_ATLAS,chargino_viastau_CMS}.}.

\subsection{Low-M Region: $M\lesssim10^7 \gev$}
 In this region, represented by benchmarks P1 and P2 in Table \ref{tab:benchmarks}, the lightest  stau is the NNLSP (with smuon/selectron co-NNLSPs)
and therefore the singlino is produced in
association with either tau or  leptons. Since the splitting between sleptons and the singlino is around 200 GeV or more, one expects high-$p_T$ taus
or leptons in the final state, which presumably can be used to reduce QCD backgrounds considerably. In this region the singlino decays (via light
pseudoscalar) to $b\bar{b}$ still inside the detector. However, identifying these displaced $b$-jets might be challenging since they are expected to
be very soft due to the small pseudoscalar mass. 
% BEN: new
We note that the low-M region is constrained (as are all GMSB models) by the
matter power spectrum as 
inferred from the Lyman$-\alpha$ forest data and WMAP~\cite{Viel:2005qj}, which disfavours
a gravitino mass between 16~eV and $m_{3/2}^{\rm crit}$, where $m_{3/2}^{\rm crit}
\sim {\cal O} ({\rm keV})$. In 
fact, from our scan we do not find any gravitinos with mass
less than 16 eV. The other bound implies that
\begin{equation}
c \tau_{\tilde{N}_1} \gtrsim 17 \, {\rm m} \, \left(
  \frac{100\textrm{~GeV}}{M_{\tilde{N}_1}} \right)^5 \left( \frac{m_{3/2}^{\rm crit}}{
      \textrm{~keV}}\right)^2 
\end{equation}
from Eqs.~(\ref{m32}),(\ref{ctau}). We remind the reader that $M_{\tilde{N}_1}$ is
close to 100 GeV because of the sum rule Eq.~(\ref{msing}). P1 and P2
violate this bound, whereas P3 may or may not, depending on the
precise value of $m_{3/2}^{\rm crit}$. However, entropy production after gravitino decoupling
may provide a cosmological evasion of the
bound for any point~\cite{Baltz:2001rq,Fujii:2002fv}.  

 \subsection{Medium-M Region: $10^7 \gev\lesssim M\lesssim10^9 \gev$}
 In this region, represented by benchmark P3, the singlino LSP is long-lived. Stau is still NNLSP, but smuon/selectron are no longer co-NNLSPs because
they are heavier than the bino-like neutralino. In consequence, a vast majority
of gluino and squark decay chains ends in stau NNLSP decaying to tau and quasi-stable singlino NLSP, with two high-$p_T$ taus in each event. 

 \subsection{Large-M Region: $M\gtrsim10^9 \gev$}
For large messenger scales, represented by benchmarks P4 and P5, the NNLSP is bino-like. Therefore the (quasi-stable) singlino is typically produced
in association with the 125 GeV Higgs, BR$(\tilde{N}_2\to\tilde{N}_1 h_2)\sim70-75\%$, or the singlet-like
Higgs, BR$(\tilde{N}_2\to\tilde{N}_1 h_1)\sim25-30\%$. Both Higgs states decay dominantly to $b \bar b$. Using a $b-$jet tagging efficiency of 70$\%$
\cite{bjettaggingCMS},  one still expects in each event at least two (three) identified high-$p_T$ $b-$jets from bino decays with a probability of
about 60 (30)$\%$. 
This comes on top of the $b-$jets originating from other decays in the gluino and/or squark decay chains. Therefore, this model can be easily
discriminated against MSSM models using searches with large numbers of $b-$jets.  

\section{Summary and Conclusions}
In this letter we have re-analyzed the DGS model, a simple and predictive framework for combining MGM and the NMSSM. We have found new regions in the
parameter space with a singlet at ${\cal O}(90-100) \gev$ and singlet-Higgs mixing giving substantial contributions to the tree-level Higgs mass.
While these regions are compatible with Higgs precision data, we find that colored sparticles can be close to their direct search limits, in sharp
contrast to MGM.  

SUSY decays have more visible particles and can lead to less missing $E_T$ as compared to MSSM predictions. The 
phenomenology is controlled by a single parameter, which determines whether SUSY decays chains lead to additional $b-$jets or taus and involve
displaced vertices. As the production cross-sections of colored sparticles are ${\cal O}(10-100)$ fb, significant parts of parameter space are
discoverable in the first year of LHC Run II.

\noindent {\bf Acknowledgements. }  
This work has been partially supported by STFC grant ST/L000385/1. BCA and MB
thank the Cambridge supersymmetry working group for helpful discussions. We thank L.~Calibbi for discussions and collaboration at an early stage of
this project. RZ thanks A.~Falkowski, M.~Goodsell, D.~Redigolo and in particular P.~Slavich for useful discussions.  This work made in the ILP LABEX
(under reference ANR-10-LABX-63) was partially supported by French state funds managed by the ANR within the Investissements d'Avenir
programme under reference ANR-11-IDEX-0004-02. The authors acknowledge the support of France Grilles for providing computing resources on the French
National
Grid Infrastructure.
This work is a part of the ``Implications of the Higgs boson discovery on supersymmetric extensions of the Standard Model'' project funded within the
HOMING
PLUS programme of the Foundation for Polish Science. MB has been partially supported by National Science Centre under research grant
DEC-2012/05/B/ST2/02597 and
the MNiSW grant IP2012 030272.

\end{document}